# Ultrahigh room-temperature hole conductivity in a perovskite cuprate with vanishing electron-correlation

Meng Wang[1,2]*†, Jianbing Zhang[3]†, Liang Si[4,5]†, Sijie Wu[3], Caiyong Li[3], Wenfeng Wu[5,6,7], Xiaodong Zhang[4], Cong Li[3], Lu Wang[8], Fachao Li[1], Lingzhi Wen[3], Yang Liu[3], Jinling Zhou[3], Masahiro Sawada[9], Nianpeng Lu[8], Qing He[10], Peng Gao[11], Tian Liang[3,12], Shuyun Zhou[3,12], Yeliang Wang[1]*, Fumitaka Kagawa[2,13], Pu Yu[3,12]*

[1]School of Integrated Circuits and Electronics, MIIT Key Laboratory for Low-Dimensional Quantum Structure and Devices, Beijing Institute of Technology; Beijing 100081, China.

[2]RIKEN Center for Emergent Matter Science (CEMS); Wako 351-0198, Japan.

[3]State Key Laboratory of Low Dimensional Quantum Physics and Department of Physics, Tsinghua University; Beijing 100084, China.

[4]School of Physics and Shaanxi Key Laboratory for Theoretical Physics Frontiers, Northwest University; Xi'an 710127, China.

[5]Institute of Solid State Physics, TU Wien; 1040 Vienna, Austria.

[6]Key Laboratory of Materials Physics, Institute of Solid State Physics, HFIPS, Chinese Academy of Sciences; Hefei 230031, China.

[7]Science Island Branch of Graduate School, University of Science and Technology of China; Hefei 230026, China.

[8]Beijing National Laboratory for Condensed Matter Physics, Institute of Physics, Chinese Academy of Sciences; Beijing 100190, China.

[9]Hiroshima Synchrotron Radiation Center, Hiroshima University; Higashihiroshima, Hiroshima 739-0046, Japan.

[10]Department of Physics, Durham University; Durham DH13LE, UK.

[11]Electron Microscopy Laboratory, School of Physics, Peking University; Beijing 100871, China.

[12]Frontier Science Center for Quantum Information; Beijing 100084, China.

[13]Department of Physics, Tokyo Institute of Technology; Tokyo 152-8551, Japan.

†These authors contributed equally to this work.

*Corresponding authors. Emails: wangmeng@bit.edu.cn (M.W.), yeliang.wang@bit.edu.cn (Y.W.) and yupu@tsinghua.edu.cn (P.Y.)

**Abstract:** Electron-correlated two-dimensional (2D) cuprates have been extensively studied since the discovery of high-$T_c$ superconductivity, in contrast, the three-dimensional (3D) counterpart perovskite cuprates remain largely unexplored due to their chemical instability and synthesis challenges. Herein, we develop an efficient two-step approach that combines symmetry-selective growth and topotactic oxidization to synthesize high-quality perovskite $LaCuO_3$ films, and furthermore reveal its exotic electronic states. The compressively strained $LaCuO_3$ films exhibit an unexpected ultrahigh *p*-type conductivity of ~$1.5\times10^5$ $\Omega^{-1}\cdot cm^{-1}$ with a hole mobility of ~30 $cm^2\cdot V^{-1}\cdot s^{-1}$ at room-temperature. X-ray absorption spectra and first-principles calculations unveil a ligand-hole state of *p-d* hybridization with degenerate $e_g$ orbitals and light effective mass, indicating nearly-vanishing electron-correlation. These features contrast sharply with 2D cuprates and offer physical insights into the design of high-performance electronic devices.

**Main text:** Dimensionality control is an important knob to manipulate the interactions of charge, spin and orbital degrees of freedom in transition metal oxides, generating diverse quantum states *(1–5)*. 2D layered cuprates have long served as a fertile model system in the studies of strong electron correlation, high-$T_C$ superconductivity, and novel spin textures *(6–10)*. While it is natural to extend the study by considering what unique properties the 3D counterparts might possess, the lack of high-quality materials has extremely hindered the researches of the emergent phenomena in 3D copper-based perovskites and corresponding artificial structures. In contrast to the extensively studied 2D cuprates *(7–9)*, such as the Ruddlesden-Popper (R-P) phase (La$_{1-x}$Sr$_x$)$_2$CuO$_4$ (**Fig. 1A**), the 3D perovskite cuprates (**Fig. 1B**) remain far from being thoroughly studied. The main reason behind this is that the perovskite ACuO$_3$ system (A = La or Sr) has a high energy Cu$^{3+}$ or Cu$^{4+}$ valence state, which prefers to convert into a more stable Cu$^{2+}$ state, driving a grand challenge for material synthesis *(11–18)*.

Starting from the 1980s, efforts have been made to fabricate LaCuO$_3$ in the pursuit of new high-$T_C$ superconducting materials and the exploration of their underlying mechanisms. Nevertheless, the reported materials are predominantly polycrystalline, and oxygen vacancies prove extremely difficult to eliminate, even when subjected to ultrahigh synthesis pressures *(11–17)*. Some of the studies suggest the LaCuO$_3$ to be a nonmagnetic bad metal *(14–17)*, while others suggest a weak-ferromagnetic bad metal *(11–13)* or an antiferromagnetic insulator *(18)*. The controversy can be mainly attributed to the inevitable grain boundaries, the competing phase, and oxygen vacancies in these polycrystalline samples *(11–18)*, which likely introduce significant extrinsic contributions to the electronic and magnetic states of perovskite cuprates.

Thus, fabricating high-quality LaCuO$_3$ sample is the key for unveiling/manipulating its intrinsic properties and unlocking potential applications.

One promising approach for fabricating stoichiometric LaCuO$_3$ thin-film involves using off-stoichiometric LaCuO$_{2.5}$ as a precursor, where Cu$^{2+}$ serve as an important intermediate state, and subsequently inserting oxygen ions via a topotactic phase transformation *(15, 19–21)*. LaCuO$_{2.5}$ has been studied for its spin ladder structure due to ordered oxygen vacancies. However, bulk synthesis of LaCuO$_{2.5}$ is challenge due to its ultra-narrow growth window (1050 ± 25 °C) and the presence of competing La$_2$CuO$_4$ phase *(22, 23)*. Previous attempts to grow LaCuO$_{2.5}$ thin film on SrTiO$_3$ (001) substrates resulted in poor-quality polycrystalline grains, highlighting the difficulty of achieving its pure phase *(24)*.

In this work, we address the challenge of synthesizing high-quality LaCuO$_3$ films by leveraging the distinct crystalline symmetries of these two competing phases, La$_2$CuO$_4$ and perovskite-like LaCuO$_{2.5}$. La$_2$CuO$_4$ exhibits a layered structure with lattice constants characterized by $a = b \ll c$ *(7–9)*, whereas LaCuO$_{2.5}$ displays a cubic-like symmetry with $a \approx b \approx c$ *(25)*. To selectively promote the growth of LaCuO$_{2.5}$ and suppress the formation of La$_2$CuO$_4$, we propose using perovskite oxide substrates with a (110) facet rather than the conventional (001) orientation, as illustrated in **Fig. 1, C and D**. This substrate choice facilitates the preferential growth of the perovskite-like phase along [110] axis. Utilizing this approach, we successfully fabricated stoichiometric LaCuO$_3$ films on a series of strained substrates via a topotactic phase transition process (i.e., ozone annealing) (**Fig. 1, E and F**). The obtained LaCuO$_3$ films exhibit a paramagnetic metallic behavior. Notably, the compressively strained LaCuO$_3$/LaAlO$_3$ (LCO/LAO) film achieves an ultrahigh conductivity of ~1.5 × 10$^5$ Ω$^{-1}$·cm$^{-1}$ and a hole-type carrier mobility of ~30 cm$^2$·V$^{-1}$·s$^{-1}$ at room-temperature. These values represent the highest reported room-temperature hole-type conductivity in known materials, significantly challenging the conventional notion of poor conductivity in correlated cuprates. Our investigations using X-ray linear dichroism (XLD) spectra further reveal that the $e_g$ orbitals in LaCuO$_3$ films are nearly degenerate, indicating a suppressed crystal-field and Jahn-Teller splitting. This is in stark contrast to the electronic states of 2D cuprates, which show distinctly-splitting $e_g$ orbitals with strong correlation. First-principles calculations elucidate that the high carrier density and mobility in LaCuO$_3$ can be attributed to nearly uncorrelated ligand-hole states, resulting from the hybridization between oxygen $p$-orbitals and copper $d$-orbitals. This unique electronic configuration underpins the exceptional transport properties observed in the perovskite LaCuO$_3$ films.

**Perovskite LaCuO$_3$ thin film fabrications**

LaCuO$_{2.5}$ thin films were first grown by pulsed laser deposition (PLD) on both SrTiO$_3$ (110) and (001) substrates for comparison (see **Methods**). Representative X-ray diffraction (XRD) results are presented in **fig. S1**. **Figure 1, G and H** exhibit the growth phase diagrams of La-Cu-O components deposited on (001) and (110) facets, respectively. The results demonstrate that the (110) substrate dramatically promotes the formation of the pure LaCuO$_{2.5}$ phase across a broad regime, a feat unattainable on (001) substrates.

The topotactic oxidization process was conducted in an ozone oven (see **Methods**). As shown in **Fig. 1I**, the XRD 2$\theta$–$\omega$ curve of pristine LaCuO$_{2.5}$ film exhibits a distinct superlattice peak at ~17°, consistent with the ordered oxygen-vacancies in the perovskite-like structure (**Fig. 1E**) *(25)*. After annealing the film at 300 °C in ozone for half an hour, the superlattice peak disappeared entirely, accompanied by a shift of the pseudocubic (110) peak from 34.2° to 33.1°, implying the formation of a pure perovskite LaCuO$_3$ phase *(11–17)*. *Ex-situ* annealing experiments in ozone environment revealed a critical phase transformation temperature of ~250 °C (**fig. S2A**). Besides, *in-situ* annealing of LaCuO$_3$ in air demonstrated a phase-stable temperature of 200 °C, above which a reversible phase transformation back to LaCuO$_{2.5}$ occurs (**fig. S2, B and C**). Notably, the oxidation process induced a dramatic color change (**Fig. 1J**). Optical transmittance measurements showed that LaCuO$_{2.5}$ is transparent in both the visible-light and infrared regions, whereas LaCuO$_3$ becomes opaque in these regions (**fig. S2D**).

To verify the formation of LaCuO$_3$ phase, we conducted both soft X-ray absorption spectroscopy (XAS) and scanning transmission electron microscopy (STEM) measurements (see **Methods**), which provide critical insight into the chemical valence state and atomic structure, respectively. As shown in **Fig. 2A**, The $L_3$-absorption peak of copper in the ozone-annealed film exhibited a blue-shift of approximately 0.6 eV compared to that in the as-grown LaCuO$_{2.5}$ film. Additionally, a distinct hump emerged between the $L_3$ and $L_2$ peaks (indicated by the shaded area), consistently suggesting the formation of Cu$^{3+}$ *(26–27)*. **Figure 2B** shows the variation of the oxygen $K$-edge, with a red-shift and enhancement of the pre-absorption peaks upon ozone annealing, suggesting an enhanced orbital hybridization between the copper $d$-orbital and the oxygen $p$-orbital *(19, 26)*. Cross-section annular bright field (ABF) images of the two phases are presented in **Fig. 2, C and D**. The La/Cu ions in the LaCuO$_{2.5}$ film show a zig-zag stacking along the out-of-plane direction, marked by the blue line. A meticulous analysis of the ionic intensity along this chain revealed a double periodicity in the lattice (**Fig. 2E**), consistent with the characteristics of LaCuO$_{2.5}$ *(25)*. In contrast, the ozone-annealed sample displayed a perfect perovskite stacking periodicity, confirming the formation of the LaCuO$_3$ phase *(19)*. High-angle annular dark field (HAADF) images also supported these findings (**fig. S3**). Energy dispersive X-ray spectra (EDS) measurements showed a sharp interface between the film and the substrate (**fig. S4**).

**Strain manipulations of LaCuO$_3$ and emergent ultrahigh conductivity**

To clarify the intrinsic electronic states of LaCuO$_3$, we conducted electrical transport and magnetization measurements on a set of high-quality epitaxial thin films with different strain states (See **Methods**). The LaCuO$_3$ film grown on SrTiO$_3$ exhibits a small biaxial tensile strain of 0.8 % (**Fig. 3A**). The temperature-dependent resistivity ($\rho$–$T$) curve exhibits a distinct metallic behavior, with a room-temperature resistivity four orders of magnitude lower than that of the insulating LaCuO$_{2.5}$ (**Fig. 3B**). This significant reduction in resistivity is consistent with the oxidization-induced color change observed in **Fig. 1J**. The magnetization of LaCuO$_3$ showed a linear dependence on the magnetic field at 10 K, with an extremely low magnetic susceptibility (**fig. S5A**), ruling out any ferromagnetic behavior. The temperature dependent magnetization curve (*M–T*) shows a combination of two components: one is proportional to $T^{-1}$ generally indicating a contribution from defects, and the other is *T*-insensitive indicating a Pauli paramagnetism in metals (**fig. S5B**) *(28)*. These features, collectively, demonstrate that the LaCuO$_3$ film has a paramagnetic ground state.

Note that epitaxial strain is a powerful means of manipulation in oxide thin films *(4, 29–31)*. It is interesting to explore how epitaxial strain can tune the electronic states of LaCuO$_3$. In this work, the (LaAlO$_3$)$_{0.3}$–(SrAl$_{0.5}$Ta$_{0.5}$O$_3$)$_{0.7}$ (LSAT) and LaAlO$_3$ (LAO) substrates were used to achieve compressive strains of –0.2 % and –1.3 %, respectively. Reciprocal space mappings (RSMs) suggested that all LaCuO$_3$ films have coherent in-plane registration with their substrates (**Fig. 3A**). Out-of-plane (OOP) lattice measurements using XRD yield a systematically increasing lattice constant along the [110] direction with increasing compressive strain (**fig. S6A**). The $\rho$–$T$ curve revealed that the room-temperature resistivity of LaCuO$_3$ is further reduced by compressive strains (**Fig. 3B**), and the film grown on LaAlO$_3$ exhibits a slightly higher residual resistivity at low temperatures compared to that on LSAT, likely due to twin boundaries from the LaAlO$_3$ substrate.

All LaCuO$_3$ films shows positive magnetoresistance with a parabolic-like dependence on the magnetic field (**fig. S7, A–C**), consistent with paramagnetic metallic behavior. Hall resistivity ($\rho_{yx}$) measurements (**fig. S7, D–F**) shows that the LaCuO$_3$ film on SrTiO$_3$ exhibits a two-carrier-types' response at low temperatures *(32)*, transitioning to a hole-type response above 150 K. The film on LSAT shows a slightly enhanced hole-type contribution (**fig. S7E**). In contrast, the film grown on LaAlO$_3$ exhibits a purely hole-type response from 2 K to 300 K (**fig. S7F and Fig. 3C**). XAS comparisons among the three samples reveal a slight blue-shift in the copper $L_3$-edge from SrTiO$_3$ to LaAlO$_3$, suggesting an enhanced Cu$^{3+}$ contribution

*(26)*. This can be attributed to the correlation between oxygen vacancy formation energy and epitaxial strain, in which the tensile (compressive) strain favors (disfavors) the formation of oxygen vacancy *(33)*. Besides, the compressive strain can further enhance the *p-d* hybridization. Both of these effects facilitate the transition to a hole-type state, in particular, the holes in O-*p* orbitals are induced by *p-d* hybridization. **Figure 3D** summarizes the temperature-dependent carrier mobility of the three films. The compressive strain effectively enhances the hole mobility, raising the room-temperature mobility from 5 cm$^2$ V$^{-1}$ s$^{-1}$ in SrTiO$_3$ to 34 cm$^2$ V$^{-1}$ s$^{-1}$ in LaAlO$_3$. Notably, the room-temperature hole mobility of the LaCuO$_3$/LaAlO$_3$ film exceeds that of the two-dimensional hole gas (2DHG) at the interfaces of LaAlO$_3$/SrTiO$_3$ (~1 cm$^2$·V$^{-1}$·s$^{-1}$) *(34)* and GaN/AlN (~20 cm$^2$·V$^{-1}$·s$^{-1}$) *(35)*, despite its relatively high carrier density (**Fig. 4A**).

The combination of high mobility and high carrier density results in an ultrahigh *p*-type room-temperature conductivity in the compressively strained LaCuO$_3$ film (**Fig. 4A**). High-conductivity materials are crucial for optimizing the metal-semiconductor contacts, reducing voltage drops, and minimizing power dissipation in electronic devices. However, as shown in **Fig. 4B**, most of the materials with high room-temperature conductivity are of the *n*-type, while *p*-type conducting materials are scarce *(36–41)*. The LaCuO$_3$ film grown on LaAlO$_3$ substrate exhibits the highest room-temperature hole conductivity ( ~1.5 × 10$^5$ Ω$^{-1}$·cm$^{-1}$) among the known materials *(7, 42–44)*, surpassing even topological materials like NbP *(32, 45)*. It has been very close to that of the *n*-type conductor copper ( ~5 × 10$^5$ Ω$^{-1}$·cm$^{-1}$) *(39)*. This result contrasts sharply with previously reported poor metallic behavior in poly-crystalline cuprate samples *(11–17)*, unveiling the intrinsic properties of LaCuO$_3$ and highlighting the profound effects of strain engineering.

**Orbital information and first-principles calculations of LaCuO$_3$**

3*d* oxides, especially cuprates, usually host relatively poor conductivity due to their narrow bandwidth, strong electron correlation, and consequently heavy effective mass. For example, the room-temperature conductivity of the R-P phase La$_{2-x}$Sr$_x$CuO$_4$ is more than two orders of magnitude lower than that of the LaCuO$_3$ films (**Fig. 4B**) *(7)*, despite both materials featuring a *p-d* hybridized Fermi surface. To elucidate the underlying orbital differences between these two structures (**Fig. 5, A and B**), we conducted XLD measurements in total fluorescence yield (FY) mode (See **Methods**). In the R-P phase La$_{2-x}$Sr$_x$CuO$_4$ system, the CuO$_6$ octahedra are elongated along the *c*-axis, resulting in the reduced energy level for $d_{z^2}$ orbital and a pronounced splitting of the $e_g$ orbitals (**Fig. 5A**) *(7, 8)*. This splitting gives rise to a significant XLD intensity at the copper *L*-edges *(46)*. Similar $e_g$ splitting can also be observed in the Y-Ba-Cu-O and LaCuO$_{2.5}$ systems with CuO$_5$ pyramid structure *(46)*. In contrast, the perovskite structure of LaCuO$_3$ exhibits a regular CuO$_6$ octahedra with negligible structural and lattice deformation, leading to the nearly degenerate $e_g$ orbital (**Fig. 5B**). The experimental setup for XLD measurements on LaCuO$_3$ films is schematically illustrated in the inset of **Fig. 5C**. By aligning the polarization of the incident light parallel to the *a*-axis and *c*-axis of the perovskite structure, we probed the differences between the $d_{z^2}$ and $d_{x^2-y^2}$ orbitals. Strikingly, the XLD intensities of LaCuO$_3$ films are dramatically reduced as compared to those of LaCuO$_{2.5}$ and La$_{2-x}$Sr$_x$CuO$_4$ (**Fig. 5C**). This observation suggests that the $e_g$ orbital splitting in LaCuO$_3$ is negligibly small (**Fig. 5B**), which could account for the high hole density and high mobility in this system.

To correlate with our experimental observation and further investigate the electronic states of LaCuO$_3$ film, we performed first-principles density-functional theory and dynamical mean-field theory calculations (DFT+DMFT) (for computational details see **Methods**). **Figure 5, D and E** show the partial density of states (PDOS) profiles of the copper *d*-orbitals and oxygen *p*-orbitals and spectral functions (DMFT-level bands), obtained from dynamical mean-field theory (DMFT) calculations for LaCuO$_3$ with the lattice grown on LaAlO$_3$. The projected density of states (PDOS) analysis reveals that each oxygen atom within the CuO$_6$ octahedron possesses a *p*-orbital that hybridizes with the Cu-$e_g$ orbitals, and it is consistent with the XLD that the two $e_g$ orbitals are degenerate (**Fig. 5D**). This hybridization results in both the Cu-$e_g$ and O-$p$ orbitals crossing the Fermi level and facilitates electron transfer from the O-$p$ orbitals to the copper orbitals, leading to the formation of ligand-hole state. The $p$-$d$ hybridization and the resulting self-doping effect leads to a heavily doped two-band $e_g$ system, which suppresses strong electron correlations. This suppression is evidenced by the similarity between DMFT and DFT band, as well as the absence of pronounced higher- and lower- Hubbard bands (**fig. S8** and **Fig. 5E**). These results suggest that the $e_g$ electrons are less localized and more itinerant, diminishing the prominence of correlation-induced features typically associated with Hubbard bands *(47)*. As a consequence, the effective mass of the $e_g$-orbital holes computed by DMFT is approximately 1.28 $m_e$ (mass of free electrons), remarkably close to non-interacting electrons. These ligand-hole states facilitate enhanced charge transport, leading to increased hole mobility. In addition, the symmetric profiles of the spin-up and spin-down PDOS indicate a paramagnetic state (**Fig. 5D**), aligning with the above experimental observations. These findings underscore the intricate interplay between Cu-$e_g$ and O-$p$ orbitals in determining the electronic properties of LaCuO$_3$ films.

**Discussion**

In this work, we present an effective strategy for synthesizing chemically metastable LaCuO$_3$ films and reveal an ultrahigh room-temperature hole conductivity in compressively strained thin films. The discovery of ultrahigh room-temperature hole conductivity and high mobility in LaCuO$_3$ films breaks the long-held perception of strong electron-correlation in cuprates, and offers physical insights into the design of high-performance oxide electronic devices. Moreover, the current study paves a promising way for the study of emergent phenomena in electron/hole- doped perovskite cuprates and their artificial superlattices, as well as a wide array of other metastable perovskite oxides (e.g., BiNiO$_3$) *(48–50)*.

**Acknowledgments:** The authors acknowledge experimental assistance from Xiaoxue

Chang, Yuanwei Sun, and Ze Hua. This study was financially supported by the National Natural Science Foundation of China (NSFC Grant Nos. 52025024, 52388201 and 12421004), and the National Key R&D program of China (Grant No. 2023YFA1406400 and 2021YFA1400100); J. Zhang acknowledges support from the China Postdoctoral Science Foundation (grant No. 2024M761596) and the Postdoctoral Fellowship Program of CPSF (grant No. GZB20240380). L.S. acknowledges the funding from the National Natural Science Foundation of China (Grant No. 12422407). L.S. also acknowledges funding through the Austrian Science Funds (FWF) project I 5398. Calculations have been mainly done on the Vienna Scientific Cluster (VSC) and National Supercomputing Center in Northwest University, Xi'an.

**Author contributions:** M. W., Y. W. and P. Y. conceived the research. M. W. and J. Zhang grew the films and performed the XRD measurements. M. W. performed the transport measurements with the help of L. Wen, T. L. and F. K. J. Zhang, Caiyong L. and F. L. performed the magnetic measurements. L. S., W. W. and X. Z. performed the first-principles calculations. M. W., S. W., M. S. and Q. H. performed the XAS and XLD measurements. J. Zhang, Y. L., Cong L., and J. Zhou performed the optical measurements. L. Wang, and N. L. performed the *in-situ* XRD measurements. P. G. and Y. W. supported the STEM measurements. M. W., J. Zhang, L. S., Y. W. and P. Y. wrote the manuscript, and all authors discussed the results and commented on the manuscript.

**Competing interests:** The authors declare no competing financial interests.

**Data and materials availability**: The data supporting the findings of this study are available from the corresponding authors upon reasonable request.

**Supplementary Materials**

Materials and Methods

Figs. S1 to S8

References (*51–65*)

**Figures and captions**

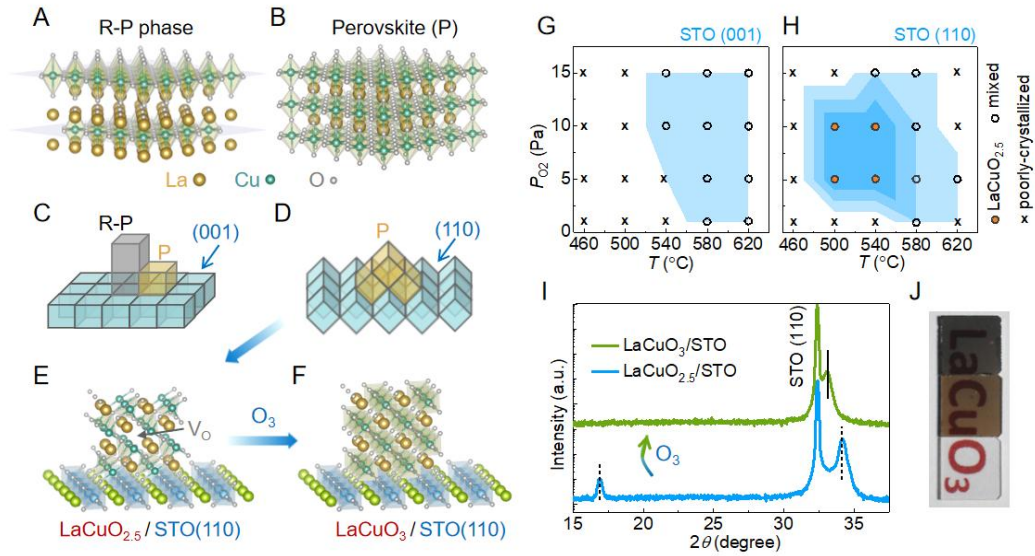

**Fig. 1 | Synthesis strategy to achieve perovskite LaCuO₃ films. (A, B)** Crystal structures of the Ruddlesden-Popper (R-P) phase $La_2CuO_4$ (**A**) and perovskite phase $LaCuO_3$ (**B**). **(C, D)** Schematic illustrations of phase control by the facet-symmetry of $SrTiO_3$ (STO) substrates. STO (001) has a square in-plane lattice symmetry (**C**), which indiscriminately supports the growth of both metastable perovskite-like (P) $LaCuO_{2.5}$ and the competing R-P phase $La_2CuO_4$. In contrast, the STO (110) facet has a rectangle in-plane lattice (**D**), which dramatically suppresses the growth of R-P phase due to a distinct lattice mismatching. **(E, F)** Schematic topotactic phase transition from $LaCuO_{2.5}$ (**E**) to $LaCuO_3$ (**F**) via an ozone-annealing process. The gray arrow in (**E**) indicates the oxygen vacancy ($V_O$). **(G, H)** Growth phase diagrams of La-Cu-O components deposited on STO (001) (**G**), and STO (110) (**H**) substrates, respectively. Both temperature and growth oxygen pressure were systematically mapped. The single phase $LaCuO_{2.5}$, mixed-phase, and poorly-crystallized states were distinguished by filled circles, open circles, and crosses, respectively. **(I)**, XRD $2\theta-\omega$ curves of $LaCuO_{2.5}$ and $LaCuO_3$ films grown on STO (110). **(J)**, Photographs of $LaCuO_3$, $LaCuO_{2.5}$ and the bare STO substrate (from top to bottom). Film thickness, 60 nm. a.u., arbitrary unit.

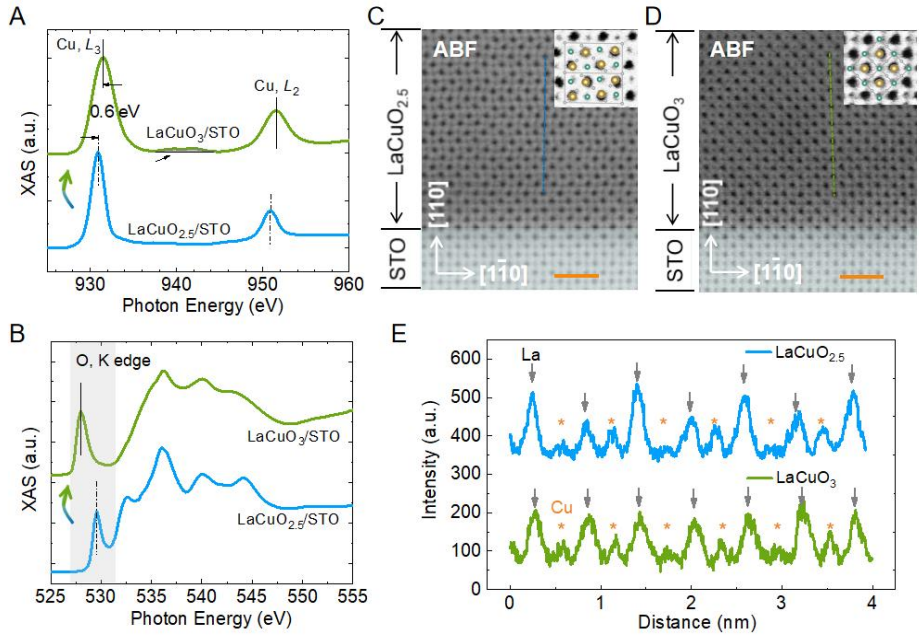

**Fig. 2 | Topotactic phase transformation from LaCuO$_{2.5}$ to LaCuO$_3$. (A, B)** X-ray absorption spectra (XAS) of copper *L*-edges (**A**), and oxygen *K*-edges (**B**) measured in LaCuO$_{2.5}$ and LaCuO$_3$ films grown on STO (110) substrate. The shifts of the *L*$_3$- and *L*$_2$-peak positions toward high energy, along with the emerged humps marked by the gray area, collectively indicate an increased valence state of copper from +2 to +3. The red-shift and enhancement of the pre-absorption peaks of oxygen *K*-edges from LaCuO$_{2.5}$ to LaCuO$_3$ correspond to an enhanced *p-d* hybridization due to the oxidization process. **(C, D)** Cross-section annular bright field (ABF) images of the LaCuO$_{2.5}$/STO (**C**) and LaCuO$_3$/STO (**D**) films. The zone axis is [001], and the orange scale bar represents 1 nm. Insets show that the ABF images match well with the calculated lattice structures of LaCuO$_{2.5}$ and LaCuO$_3$, respectively. **(E)**, Atomic intensity distributions along the out-of-plane directions in both LaCuO$_{2.5}$ and LaCuO$_3$, obtained from the blue and green lines in (**C**) and (**D**). Arrows and stars indicate La and Cu ions, respectively. LaCuO$_{2.5}$ shows a distinct period-doubling due to the zigzag chains of La/Cu ions along the [110] direction, in contrast to the straight chains in the perovskite LaCuO$_3$.

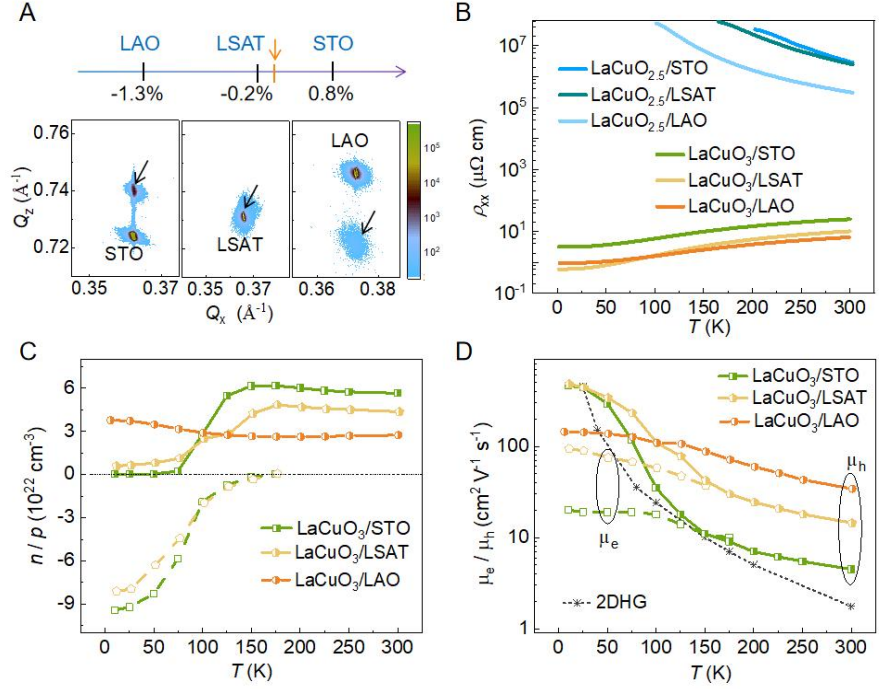

**Fig. 3 | Transport properties of strained LaCuO₃ films. (A)**, Illustration of the epitaxial strain by different substrates, and the corresponding reciprocal space mappings (RSMs) of LaCuO$_3$ films grown on STO, LSAT, and LAO (110) substrates. The lattice constant of bulk LaCuO$_3$ (pseudo-cubic) is indicated by the yellow arrow. **(B)**, Temperature-dependent resistivity ($\rho_{xx}$–$T$) curves of the LaCuO$_{2.5}$ and LaCuO$_3$ films grown on STO, LSAT and LAO (110) substrates. The film thickness is 60 nm. **(C, D)** Temperature-dependent carrier density **(C)**, and mobility **(D)** of LaCuO$_3$ films grown on three different substrates. For comparison, the hole mobility of a two-dimensional hole gas (2DHG) at the SrTiO$_3$/LaAlO$_3$ (STO/LAO) interface is shown *(34)*.

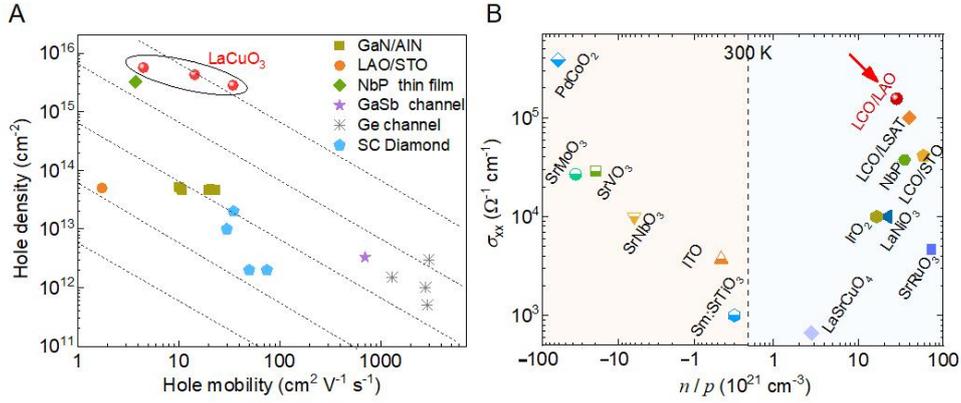

**Fig. 4 | Comparison of the room-temperature transport properties**. (**A**) Comparison of room temperature hole density and mobility among LaCuO$_3$ films and several representative high hole-mobility conductors. The materials include the 2DHG at the interfaces of LAO/STO and GaN/AlN *(34, 35)*, 2DHG at the surface-conducting diamond (SC Diamond), GaSb-channel, and Ge-channel *(34)*, and conducting NbP thin film *(45)*. The 2D hole density for LaCuO$_3$ and NbP was calculated assuming a thickness of 1 nm. Dash lines extending from the bottom-left to the top-right signify an increase in conductivity by one order of magnitude. (**B**) Room temperature conductivity ($\sigma_{xx}$) and the corresponding carrier density for LaCuO$_3$ films compared with a series of representative metallic transition metal oxides *(7, 36-44)*. The LaCuO$_3$/LaAlO$_3$ (LCO/LAO) film shows the highest room-temperature conductivity of hole-type carrier, as indicated by the red arrow. LCO/STO and LCO/LSAT denote the LaCuO$_3$ films grown on SrTiO$_3$ and LSAT, respectively. The topological material NbP film is also shown for comparison *(45)*.

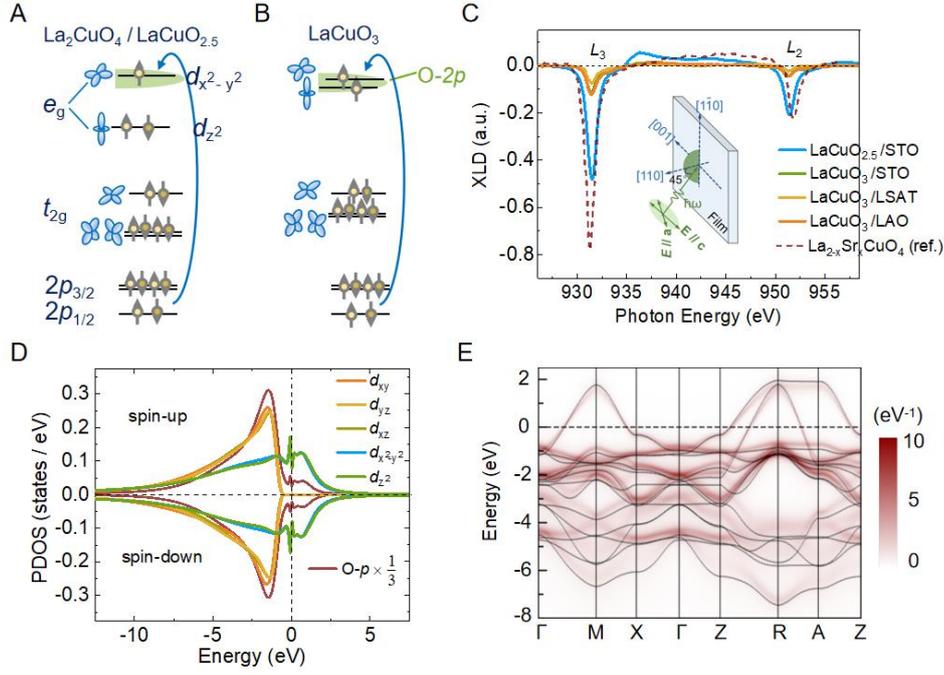

**Fig. 5 | Electronic structure of perovskite LaCuO$_3$ films. (A, B)** Illustrations of copper *d*-orbital splitting in the structures of LaCuO$_{2.5}$ or R-P phase La$_2$CuO$_4$ (**A**), and perovskite LaCuO$_3$ (**B**). (**C**) X-ray linear dichroism (XLD) spectra of the copper *L*-edges. Compared with LaCuO$_{2.5}$ and La$_{2-x}$Sr$_x$CuO$_4$ *(7, 46)*, all LaCuO$_3$ films show a dramatically reduced XLD intensity. Inset: experimental setup for XLD measurements on La-Cu-O films grown along the [110] direction, with the incident light in the (001) plane at a 45° angle to the normal direction. The polarization direction (*E*) of X-ray was controlled between parallel to the *a*-axis and *c*-axis for LaCuO$_3$ and the pseudocubic LaCuO$_{2.5}$ films, respectively. (**D**), Partial density of states (PDOS) of the copper *d*-orbitals and oxygen *p*-orbitals in the LaCuO$_3$ film, obtained by dynamical mean-field theory (DMFT) calculations using the lattice constants of LaCuO$_3$ film grown on LaAlO$_3$. The degeneracy of copper *d*-orbitals and a *p-d* hybridized Fermi-energy are manifested. (**E**), The DMFT (color spectra) and DFT (grey lines) band structures for LaCuO$_3$. The close alignment of the DFT and DMFT bands, coupled with the absence of higher- and lower-Hubbard bands, suggests that the $e_g$ electrons in this system are less localized and exhibit weak correlation effects.